\documentclass[singlecolumn]{revtex4-1}

\usepackage{tabularx}
\usepackage{array}
\usepackage{mathrsfs}
\usepackage{amssymb}
\usepackage{amsmath}
\usepackage{graphicx}
\usepackage{bm}

\usepackage{float}
\usepackage{subfigure}

\usepackage[colorlinks,
linkcolor=blue,
anchorcolor=blue,
citecolor=blue
]{hyperref}

\begin{document}
\title{Reference-frame-independent measurement-device-independent quantum key distribution with imperfect sources}

\author{Jian-Rong Zhu$^{1,2,3}$, Chun-Mei Zhang $^{1,2,3}$ \footnote{cmz@njupt.edu.cn}, Qin Wang$^{1,2,3}$ \footnote{qinw@njupt.edu.cn}}

\address{$^1$ Institute of Quantum Information and Technology, Nanjing University of Posts and Telecommunications, Nanjing 210003, China\\
$^2$ Broadband Wireless Communication and Sensor Network Technology, Key Lab of Ministry of Education, NUPT, Nanjing 210003, China\\
$^3$ Telecommunication and Networks, National Engineering Research Center, NUPT, Nanjing 210003, China}

\begin{abstract}
Reference-frame-independent measurement-device-independent quantum key distribution (RFI-MDI-QKD) can remove all potential detector side-channel attacks and the requirement of real-time alignment of reference frames. However, all previous RFI-MDI-QKD implementations assume the perfect state preparation in sources, which is impractical and may lead to security loopholes. Here, we propose a RFI-MDI-QKD protocol which is robust against state preparation flaws. Comparing to the conventional six-state RFI-MDI-QKD, our scheme can be realized with only four flawed states, which improves the practical security of RFI-MDI-QKD and simplifies the experimental implementation. In addition, simulation results demonstrate that source flaws in $Z$ basis have adverse effect on the performance of RFI-MDI-QKD while the source flaws in $X$ and $Y$ bases have almost no effect. We hope that this work could provide a valuable reference for practical implementations of RFI-MDI-QKD.
\end{abstract}

\maketitle

\section{Introduction}\label{1}

Quantum key distribution (QKD) \cite{QKD1} enables two distant parties (Alice and Bob) to share a pair of random secret keys with information-theoretical security, guaranteed by the laws of quantum physics. However, the gap between the theoretical models and practical implementations may open security loopholes to the malicious eavesdroppers (Eve). In fact, various quantum hacking attacks against certain commercial and research QKD systems have been proposed and demonstrated experimentally \cite{Attack-detector2,Attack-detector3,Attack-detector5,Attack-source1,Attack-source3,Attack}. To tackle security loopholes in realistic QKD systems, measurement-device-independent QKD (MDI-QKD) \cite{MDI1,MDI2} was proposed, which is immune to all possible detector side-channel attacks. Particularly, compared with QKD networks based on traditional trusted relays \cite{QKD-field1,QKD-field4,QKD-field6},  MDI-QKD is believed to be a powerful candidate for future quantum networks \cite{MDI-exper6}. Owing to the balance between security and practicability, experimental MDI-QKD systems have been extensively demonstrated  \cite{MDI-exper1,MDI-exper2,MDI-exper3,MDI-exper4,MDI-exper7,MDI-exper8}.

However, among these demonstrations, the real-time alignment of reference frames between Alice and Bob is essential to ensure the system stability and higher secret key rate. Although the calibration of reference frames is feasible, it is time-consuming and  makes MDI-QKD systems, particularly multi-user QKD networks, complicated. Fortunately, reference-frame-independent MDI-QKD (RFI-MDI-QKD) \cite{RFI-MDI} was proposed to merge the merits of reference-frame-independent QKD \cite{RFI-QKD} and MDI-QKD, that is, it can get rid of the calibration procedure and eliminate all detector side channels. To date, RFI-MDI-QKD has attracted great theoretical and experimental interests \cite{RFI-Zhang-PRA1,RFI-Zhang-JLT1,RFI-MDI-exper1,RFI-MDI-exper2,RFI-MDI-exper3}.

Since MDI-QKD removes all attacks aimed at measurement devices, eavesdroppers may shift their target towards sources due to various rigorous source assumptions. For example, all quantum states should be prepared perfectly without flaws, while this can not be satisfactorily met with practical encoding devices, which will compromise the security of practical MDI-QKD systems. Recently, in order to relax assumptions on the encoding systems, several protocols have been proposed and demonstrated experimentally \cite{QKD-mismatch,MDI-mis1,MDI-mis2,MDI-mis3,MDI-exper-mis,MDI-exper-mis-2,Zhu:6,QKD-sourceflaw,RFI-source,BB84-exper-sourceflaw,MDI-exper-sourceflaw,RFI-exper2}.  In particular, by exploiting the rejected-data analysis, Tamaki \emph{et al.} \cite{QKD-sourceflaw} proposed a loss-tolerant method to incorporate state preparation flaws in two-dimensional Hilbert space, which has been generalized to MDI-QKD \cite{MDI-exper-sourceflaw}. However, unlike MDI-QKD,  all previous RFI-MDI-QKD implementations \cite{RFI-MDI-exper1,RFI-MDI-exper2,RFI-MDI-exper3} still assume perfect state preparations, which is not realistic in practical systems.

In this paper, inspired by the loss-tolerant idea in Ref. \cite{QKD-sourceflaw}, we proposed a RFI-MDI-QKD protocol with imperfect sources. In contrast to the original RFI-MDI-QKD \cite{RFI-MDI} implemented by six states, our protocol only needs to prepare four encoding states (that is, two states in $Z$ basis, one state in $X$ basis and one state in $Y$ basis). More importantly, the four states in our protocol can tolerate some errors of the imperfect encoding devices,  which will simplify the experimental implementation and achieve a higher security level. Simulation results show that our protocol has a pretty good performance even with state preparation flaws.

\section{Original RFI-MDI-QKD protocol}\label{2}

\begin{figure}[htbp]
	\centering
	\includegraphics[scale=0.6]{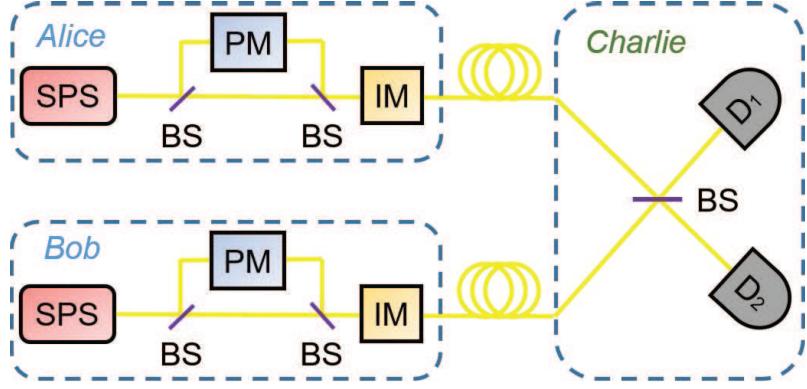}
	\caption{Schematic setup of RFI-MDI-QKD with time-bin phase encoding. SPS, single-photon source; BS, beam splitter; PM, phase modulator; IM, intensity modulator; $\;{D_j}(j = 1,2)$, single-photon detector.}
	\label{fig:Fig1}
\end{figure}

To begin with we briefly introduce the original RFI-MDI-QKD protocol with six-state encoding \cite{RFI-MDI}, where $Z$ basis can be well aligned while $X$ and $Y$ bases may drift slowly with the unknown angle ${\beta}$. Based on the setup illustrated in Fig.\ref{fig:Fig1}, Alice and Bob can randomly prepare Z-basis states $\{ \left| 0 \right\rangle ,\left| 1 \right\rangle \} $, X-basis states $\{ \left|  +  \right\rangle  = (\left| 0 \right\rangle  + \left| 1 \right\rangle )/\sqrt 2 ,\left|  -  \right\rangle  = (\left| 0 \right\rangle  - \left| 1 \right\rangle )/\sqrt 2 \} $ and Y-basis states $\{ \left| { + i} \right\rangle  = (\left| 0 \right\rangle  + i\left| 1 \right\rangle )/\sqrt 2 ,\left| { - i} \right\rangle  = (\left| 0 \right\rangle  - i\left| 1 \right\rangle )/\sqrt 2 \} $. Specifically, the $Z$-basis (time-basis) states can be prepared by restraining one of the two time bins with an intensity modulator (IM), and the $X$-basis and $Y$-basis (phase-basis) states can be prepared by modulating the relative phases between the two time bins with a phase modulator (PM). For simplicity, a single-photon source (SPS) is employed here.

After the state measurement by the third party Eve(Charlie) and some post-processing operations by Alice and Bob, the amount of information leaked to Eve is bounded by \cite{RFI-MDI}
\begin{equation}\label{IE}
	{I_E} = \left( {1 - e_{ZZ}^{11}} \right)H(1 + u/2) + e_{ZZ}^{11}H(1 + v/2),
\end{equation}
where $e_{ZZ}^{11}$ is the error rate of SPSs in $ZZ$ bases, $H(x) =  - x{\log _2}(x) - (1 - x){\log _2}(1 - x)$ is the binary Shannon entropy function, $u = \min \{ \sqrt {C/2} /(1 - e_{ZZ}^{11}),1\}$, and $v = \sqrt {C/2 - {{\left( {1 - e_{ZZ}^{11}} \right)}^2}{u^2}} /e_{ZZ}^{11}$. The quantity $C$ in $u$ and $v$ is given by $C = {(1 - 2e_{XX}^{11})^2} + {(1 - 2e_{XY}^{11})^2} + {(1 - 2e_{YX}^{11})^2} + {(1 - 2e_{YY}^{11})^2}$, where  $e_{{\alpha \chi }}^{11}$ ($\alpha ,\chi  \in \{ X,Y\} $) is the error rate of SPSs in ${\alpha \chi } $ bases.

The secret key rate of original RFI-MDI-QKD based on SPSs can be expressed as
\begin{equation}\label{R-SPS}
	R = {Q_{ZZ}^{11}}[1 - {I_E}] - {Q_{ZZ}^{11}}fH(e_{ZZ}^{11}),
\end{equation}
where ${Q_{ZZ}^{11}}$ is the gain of SPSs in $ZZ$ bases, and $f$ is the inefficiency of error correction, which is chosen as 1.16 in this work.

Notice that the original RFI-MDI-QKD \cite{RFI-MDI} strictly assumes perfect state preparation. However, the encoding apparatuses possess inherent imperfections, for instance the finite modulation accuracy of IM and PM in Fig. \ref{fig:Fig1}, and the security of real-life RFI-MDI-QKD systems is questionable. To address the problem of state preparation flaws, we propose a loss-tolerant RFI-MDI-QKD protocol with imperfect sources in Section \ref{3}.

\section{RFI-MDI-QKD with imperfect sources}\label{3}

For simplicity, we introduce our protocol based on SPSs first. In this protocol, Alice and Bob randomly prepare quantum states choosing from ${\rm{\{ }}{\hat \rho _{0Z}}{\rm{,}}{\hat \rho _{1Z}},{\hat \rho _{0X}},{\hat \rho _{0Y}}{\rm{\} }}$ , which may be different for Alice and Bob, with probability 1/4. The density matrices of their states can be expressed as a linear combination of the identity matrix ${{\hat \sigma }_I}$ and Pauli matrices $\{ {{\hat \sigma }_X},{{\hat \sigma }_Y},{{\hat \sigma }_Z}\} $ :
\begin{equation}
	\begin{aligned}
		{\hat \rho _{j\alpha' }} &= \frac{1}{2}({\hat \sigma _I} + P_X^{j\alpha' }{\hat \sigma _X} + P_Y^{j\alpha' }{\hat \sigma _Y} + P_Z^{j\alpha' }{\hat \sigma _Z}), \\ 
		{\hat \rho _{s\chi' }} &= \frac{1}{2}({\hat \sigma _I} + P_X^{s\chi' }{\hat \sigma _X} + P_Y^{s\chi' }{\hat \sigma _Y} + P_Z^{s\chi' }{\hat \sigma _Z}), 
	\end{aligned}
	\label{t md}
\end{equation}
where ${\hat \rho _{j\alpha' }}$ (${\hat \rho _{s\chi' }} $) represents the density matrix of states prepared by Alice (Bob) corresponding to the bit value $j (s) \in \{ 0,1\} $ in basis $ \alpha' (\chi')  \in \{ Z,X,Y\} $. $ P_X^{j\alpha' }$, $P_Y^{j\alpha' }$ and $P_Z^{j\alpha' } $ ($ P_X^{s\chi' }$, $P_Y^{s\chi' }$ and $P_Z^{s\chi' } $) are the coefficients of the Bloch vector of ${\hat \rho _{j\alpha' }}$ (${\hat \rho _{s\chi' }} $). And we denote ${\left| {{\phi _{j\alpha' }}} \right\rangle _{{A_e}E}}$ (${\left| {{\phi _{s\chi' }}} \right\rangle _{{B_e}E'}}$) as the purification of ${\hat \rho _{j\alpha' }}$ (${\hat \rho _{s\chi' }}$), where the subscripts ${{A_e}}$ (${{B_e}}$) and $E$ ($E'$) represent the extended system possessed by Alice (Bob) and the system sent to Eve, respectively. Then, the emission of ${\hat \rho _{jZ }}$ (${\hat \rho _{sZ }} $) can be expressed as the equivalent entanglement protocol : Alice (Bob) prepares the state $|{\Psi _Z}{\rangle _{A{A_e}E}} = ({\left| {{0_Z}} \right\rangle _A}{\left| {{\phi _{{0_Z}}}} \right\rangle _{{A_e}E}} + {\left| {{1_Z}} \right\rangle _A}{\left| {{\phi _{{1_Z}}}} \right\rangle _{{A_e}E}})/\sqrt 2$ ($|{\Psi _Z}{\rangle _{B{B_e}E'}} = ({\left| {{0_Z}} \right\rangle _B}{\left| {{\phi _{{0_Z}}}} \right\rangle _{{B_e}E'}} + {\left| {{1_Z}} \right\rangle _B}{\left| {{\phi _{{1_Z}}}} \right\rangle _{{B_e}E'}})/\sqrt 2 $) and measures system $A$ ($B$) in $Z$ basis, and sends system $E$ ($E'$) to Eve. 

Now we consider a virtual protocol that Alice (Bob) prepares the state $|{\Psi _Z}{\rangle _{A{A_e}E}}$ ($|{\Psi _Z}{\rangle _{B{B_e}E'}}$), measures system $A$ ($B$) in $\alpha$ ($\chi$) basis ($\alpha ,\chi  \in \{ X,Y\} $), and sends system $E$ ($E'$) to Eve. The virtual state sent by Alice is given as $\hat \rho _{j\alpha }^{{\rm{vir}}} = {{\mathop{\rm Tr}\nolimits} _{A{A_e}}}[\hat P({\left| {{{j\alpha }}} \right\rangle _A}) \otimes {I_{{A_e}E}}\hat P({\left| {{\Psi _Z}} \right\rangle _{A{A_e}E}})]$ \cite{QKD-sourceflaw}, where $\hat P(x) = \left| x \right\rangle \left\langle x \right|$, ${\left| {{j_X}} \right\rangle _A} = (\left| {{0_Z}} \right\rangle  + {( - 1)^j}\left| {{1_Z}} \right\rangle )/\sqrt 2 $ and ${\left| {{j_Y}} \right\rangle _A} = (\left| {{0_Z}} \right\rangle  + i{( - 1)^j}\left| {{1_Z}} \right\rangle )/\sqrt 2 $. Likewise, the virtual state sent by Bob is $\hat \rho _{s\chi }^{{\rm{vir}}} = {{\mathop{\rm Tr}\nolimits} _{B{B_e}}}[\hat P({\left| {s\chi } \right\rangle _B}) \otimes {I_{{B_e}E'}}\hat P({\left| {{\Psi _Z}} \right\rangle _{B{B_e}E'}})]$. And the normalized version of their virtual states can be expressed as
\begin{equation}
	\begin{aligned}
		\hat \rho _{j\alpha }^{{\rm{'vir}}} &= \frac{1}{2}({\hat \sigma _I} + P_X^{j\alpha ,\rm{vir}}{\hat \sigma _X} + P_Y^{j\alpha ,\rm{vir}}{\hat \sigma _Y} + P_Z^{j\alpha ,\rm{vir}}{\hat \sigma _Z}), \\ 
		\hat \rho _{s\chi }^{\rm{'vir}} &= \frac{1}{2}({\hat \sigma _I} + P_X^{s\chi ,\rm{vir}}{\hat \sigma _X} + P_Y^{s\chi ,\rm{vir}}{\hat \sigma _Y} + P_Z^{s\chi ,\rm{vir}}{\hat \sigma _Z}). 
	\end{aligned}
	\label{v pmd}
\end{equation}

Here we assume the states received by Eve are only projected into the Bell state $|{\Psi ^ - }\rangle  = (|01\rangle  - |10\rangle )/\sqrt 2 $. The error rate of SPSs $e_{{\alpha \chi }}^{11}$ in $C$ can be derived from the error rates of virtual states, and expressed as \cite{QKD-sourceflaw}
\begin{equation}\label{E}
	e_{\alpha \chi }^{11} = \frac{{Y_{0\alpha ,0\chi }^{{\Psi ^ - },{\rm{vir}}} + Y_{1\alpha ,1\chi }^{{\Psi ^ - },{\rm{vir}}}}}{{Y_{0\alpha ,0\chi }^{{\Psi ^ - },{\rm{vir}}} + Y_{0\alpha ,1\chi }^{{\Psi ^ - },{\rm{vir}}} + Y_{1\alpha ,0\chi }^{{\Psi ^ - },{\rm{vir}}} + Y_{1\alpha ,1\chi }^{{\Psi ^ - },{\rm{vir}}}}}.
\end{equation}
In Eq. (\ref{E}), $Y_{j\alpha ,s\chi }^{{\Psi ^ - },{\rm{vir}}}$ is the joint probability that Alice and Bob transmit the virtual states $\hat \rho _{j\alpha }^{{\rm{vir}}}$ and $\hat \rho _{s\chi }^{{\rm{vir}}}$ respectively and Eve declares a successful outcome $|{\Psi ^ - }\rangle$, which can be written as \cite{QKD-sourceflaw}
\begin{equation}\label{y11-vir1-T}
Y_{j\alpha ,s\chi }^{{\Psi ^ - },{\rm{vir}}}{\rm{ = Tr}}[\hat \rho _{j\alpha }^{{\rm{vir}}}]{\rm{Tr}}[\hat \rho _{s\chi }^{{\rm{vir}}}]{\rm{Tr[}}{{\rm{\hat D}}_{{\Psi ^ - }}}\hat \rho _{j\alpha }^{{\rm{'vir}}} \otimes \hat \rho _{s\chi }^{\rm{'vir}}],
\end{equation}
where ${{\rm{\hat D}}_{{\Psi ^ - }}}$ is Eve’s operation corresponding to the announcement of the outcome $|{\Psi ^ - }\rangle$, $ {\rm{Tr}}[\hat \rho _{j\alpha }^{{\rm{vir}}}]$ and $ {\rm{Tr}}[\hat \rho _{s\chi }^{{\rm{vir}}}]$ represent the probabilities of the emitted virtual states $\hat \rho _{j\alpha }^{{\rm{vir}}}$ and $\hat \rho _{s\chi }^{{\rm{vir}}}$, respectively. Define the transmission rate of ${{\hat \sigma }_l} \otimes {{\hat \sigma }_{l'}}$ ($l,l' \in \{ I,X,Y,Z\} $) as  $q_{l \otimes l'}^{{\Psi ^ - }} = \frac{1}{4}{\rm{Tr}}[{{{\rm{\hat D}}}_{{\Psi ^ - }}}{{\hat \sigma }_l} \otimes {{\hat \sigma }_{l'}}]$. Since $\hat \rho _{j\alpha }^{{\rm{'vir}}} \otimes \hat \rho _{s\chi }^{\rm{'vir}}$ can be expressed as a linear combination of ${{\hat \sigma }_l} \otimes {{\hat \sigma }_{l'}}$ ($l,l' \in \{ I,X,Y,Z\} $), Eq. (\ref{y11-vir1-T}) can be given as
\begin{equation}
	\begin{array}{l}
		Y_{j\alpha ,s\chi }^{{\Psi ^ - },{\rm{vir}}} = P_{j\alpha }^{{\rm{vir}}} P_{s\chi }^{{\rm{vir}}} (q_{I \otimes I}^{{\Psi ^ - }} + P_X^{s\chi ,{\rm{vir}}}{q_{I \otimes X}^{{\Psi ^ - }}} + P_Y^{s\chi ,{\rm{vir}}}{q_{I \otimes Y}^{{\Psi ^ - }}}  \\ \\
		+ P_Z^{s\chi ,{\rm{vir}}}{q_{I \otimes Z}^{{\Psi ^ - }}} 
		+ P_X^{j\alpha ,{\rm{vir}}}{q_{X \otimes I}^{{\Psi ^ - }}} + P_X^{j\alpha ,{\rm{vir}}}P_X^{s\chi ,{\rm{vir}}}{q_{X \otimes X}^{{\Psi ^ - }}} + \\ \\
		P_X^{j\alpha ,{\rm{vir}}}P_Y^{s\chi ,{\rm{vir}}}{q_{X \otimes Y}^{{\Psi ^ - }}} + P_X^{j\alpha ,{\rm{vir}}}P_Z^{s\chi ,{\rm{vir}}}{q_{X \otimes Z}^{{\Psi ^ - }}}+ P_Y^{j\alpha ,{\rm{vir}}}{q_{Y \otimes I}^{{\Psi ^ - }}}  \\ \\
		+ P_Y^{j\alpha ,{\rm{vir}}}P_X^{s\chi ,{\rm{vir}}}{q_{Y \otimes X}^{{\Psi ^ - }}} + P_Y^{j\alpha ,{\rm{vir}}}P_Y^{s\chi ,{\rm{vir}}}{q_{Y \otimes Y}^{{\Psi ^ - }}} + \\ \\
		P_Y^{j\alpha ,{\rm{vir}}}P_Z^{s\chi ,{\rm{vir}}}{q_{Y \otimes Z}^{{\Psi ^ - }}}
		+ P_Z^{j\alpha ,{\rm{vir}}}{q_{Z \otimes I}^{{\Psi ^ - }}} + P_Z^{j\alpha ,{\rm{vir}}}P_X^{s\chi ,{\rm{vir}}}{q_{Z \otimes X}^{{\Psi ^ - }}} \\ \\
		+ P_Z^{j\alpha ,{\rm{vir}}}P_Y^{s\chi ,{\rm{vir}}}{q_{Z \otimes Y}^{{\Psi ^ - }}} + P_Z^{j\alpha ,{\rm{vir}}}P_Z^{s\chi ,{\rm{vir}}}{q_{Z \otimes Z}^{{\Psi ^ - }}}),
	\end{array}
	\label{y11-vir1}
\end{equation}
where  $P_{j\alpha }^{{\rm{vir}}} = {\rm{Tr}}[\hat \rho _{j\alpha }^{{\rm{vir}}}]$ and $P_{s\chi }^{{\rm{vir}}} = {\rm{Tr}}[\hat \rho _{s\chi }^{{\rm{vir}}}]$. For ease of notation, Eq. (\ref{y11-vir1}) can be expressed as
\begin{equation}\label{y11-vir2}
	Y_{j\alpha ,s\chi }^{{\Psi ^ - },{\rm{vir}}} = P_{j\alpha }^{{\rm{vir}}}P_{s\chi }^{{\rm{vir}}}{\rm{P}}_{j\alpha ,s\chi }^{{\rm{vir}}}{{\rm{q}}^{\rm T}},
\end{equation}
where
\begin{equation}
	\begin{aligned}
		{\rm{P}}_{j\alpha ,s\chi }^{{\rm{vir}}} &= [1,P_X^{s\chi ,{\rm{vir}}},P_Y^{s\chi ,{\rm{vir}}},P_Z^{s\chi ,{\rm{vir}}},P_X^{j\alpha ,{\rm{vir}}},P_X^{j\alpha ,{\rm{vir}}}P_X^{s\chi ,{\rm{vir}}},\\
		& P_X^{j\alpha ,{\rm{vir}}}P_Y^{s\chi ,{\rm{vir}}},P_X^{j\alpha ,{\rm{vir}}}P_Z^{s\chi ,{\rm{vir}}},P_Y^{j\alpha ,{\rm{vir}}},P_Y^{j\alpha ,{\rm{vir}}}P_X^{s\chi ,{\rm{vir}}},\\
		& P_Y^{j\alpha ,{\rm{vir}}}P_Y^{s\chi ,{\rm{vir}}},P_Y^{j\alpha ,{\rm{vir}}}P_Z^{s\chi ,{\rm{vir}}},P_Z^{j\alpha ,{\rm{vir}}}, P_Z^{j\alpha ,{\rm{vir}}}P_X^{s\chi ,{\rm{vir}}},\\
		& P_Z^{j\alpha ,{\rm{vir}}}P_Y^{s\chi ,{\rm{vir}}},P_Z^{j\alpha ,{\rm{vir}}}P_Z^{s\chi ,{\rm{vir}}}],
	\end{aligned}
	\label{P-vir}
\end{equation}

\begin{equation}
	\begin{aligned}
		{\rm{q}} & = [q_{I \otimes I}^{{\Psi ^ - }},q_{I \otimes X}^{{\Psi ^ - }},q_{I \otimes Y}^{{\Psi ^ - }},q_{I \otimes Z}^{{\Psi ^ - }},q_{X \otimes I}^{{\Psi ^ - }},q_{X \otimes X}^{{\Psi ^ - }},q_{X \otimes Y}^{{\Psi ^ - }},q_{X \otimes Z}^{{\Psi ^ - }}, \\ 
		& q_{Y \otimes I}^{{\Psi ^ - }},q_{Y \otimes X}^{{\Psi ^ - }},q_{Y \otimes Y}^{{\Psi ^ - }},q_{Y \otimes Z}^{{\Psi ^ - }},q_{Z \otimes I}^{{\Psi ^ - }},q_{Z \otimes X}^{{\Psi ^ - }},q_{Z \otimes Y}^{{\Psi ^ - }},q_{Z \otimes Z}^{{\Psi ^ - }}],
	\end{aligned}
	\label{q}
\end{equation}
and the superscript T represents the transposition of a matrix. Once the transmission rates of the Pauli matrices ${\rm{q}}$ is obtained, we can get the values of the transmission rate $Y_{j\alpha ,s\chi }^{{\Psi ^ - },{\rm{vir}}}$ and the error rate $e_{\alpha \chi }^{11}$.

In fact, we can calculate the transmission rates of Pauli matrices q using the transmission rates of the actual states $Y_{j\alpha ',s\chi '}^{{\Psi ^ - }}$ in experiments. The transmission rates $Y_{j\alpha ',s\chi '}^{{\Psi ^ - }}$ can be expressed as
\begin{equation}\label{y11-act}
	Y_{j\alpha ',s\chi '}^{{\Psi ^ - }} = {P_{j\alpha '}}{P_{s\chi '}}{{\rm{P}}_{j\alpha ',s\chi '}}{{\rm{q}}^{\rm T}},
\end{equation}
where 
\begin{equation}
	\begin{aligned}
		{{\rm{P}}_{j\alpha ',s\chi '}} & = [1,P_X^{s\chi '},P_Y^{s\chi '},P_Z^{s\chi '},P_X^{j\alpha '},P_X^{j\alpha '}P_X^{s\chi '},P_X^{j\alpha '}P_Y^{s\chi '},\\
		& P_X^{j\alpha '}P_Z^{s\chi '},P_Y^{j\alpha '},P_Y^{j\alpha '}P_X^{s\chi '},P_Y^{j\alpha '}P_Y^{s\chi '},P_Y^{j\alpha '}P_Z^{s\chi '},\\
		& P_Z^{j\alpha '},P_Z^{j\alpha '}P_X^{s\chi '},P_Z^{j\alpha '}P_Y^{s\chi '},P_Z^{j\alpha '}P_Z^{s\chi '}],
	\end{aligned}
	\label{P-act}
\end{equation}
and ${P_{j\alpha '}}$ (${P_{s\chi '}}$) is the probability of the emitted actual states ${\hat \rho _{j\alpha' }}$ (${\hat \rho _{s\chi' }} $), for instance ${P_{j\alpha '}} = {P_{s\chi '}} = 1/4$ in this paper. Since Alice and Bob each send four states, we can construct a set of sixteen independent linear equations as
\begin{equation}
	\begin{aligned}
		Y_{jZ,sZ}^{{\Psi ^ - }} &= \frac{1}{{16}}{{\rm{P}}_{jZ,sZ}}{{\rm{q}}^{\rm T}},\\
		Y_{jZ,0X}^{{\Psi ^ - }} &= \frac{1}{{16}}{{\rm{P}}_{jZ,0X}}{{\rm{q}}^{\rm T}},\\
		Y_{jZ,0Y}^{{\Psi ^ - }} &= \frac{1}{{16}}{{\rm{P}}_{jZ,0Y}}{{\rm{q}}^{\rm T}},\\
		Y_{0X,sZ}^{{\Psi ^ - }} &= \frac{1}{{16}}{{\rm{P}}_{0X,sZ}}{{\rm{q}}^{\rm T}},\\
		Y_{0X,0X}^{{\Psi ^ - }} &= \frac{1}{{16}}{{\rm{P}}_{0X,0X}}{{\rm{q}}^{\rm T}},\\
		Y_{0X,0Y}^{{\Psi ^ - }} &= \frac{1}{{16}}{{\rm{P}}_{0X,0Y}}{{\rm{q}}^{\rm T}},\\
		Y_{0Y,sZ}^{{\Psi ^ - }} &= \frac{1}{{16}}{{\rm{P}}_{0Y,sZ}}{{\rm{q}}^{\rm T}},\\
		Y_{0Y,0X}^{{\Psi ^ - }} &= \frac{1}{{16}}{{\rm{P}}_{0Y,0X}}{{\rm{q}}^{\rm T}},\\
		Y_{0Y,0Y}^{{\Psi ^ - }} &= \frac{1}{{16}}{{\rm{P}}_{0Y,0Y}}{{\rm{q}}^{\rm T}}.\\
	\end{aligned}
	\label{equ}
\end{equation}
With the transmission rates $Y_{j\alpha ',s\chi '}^{{\Psi ^ - }}$ and ${{\rm{P}}_{j\alpha ',s\chi '}} $ obtained from a RMI-MDI-QKD experiment, we can calculate the transmission rates of the Pauli matrices $\rm{q}$. Finally, the error rate $e_{\alpha \chi }^{11}$ can be obtained with Eq. (\ref{E}) and Eq. (\ref{y11-vir2}).

Due to the shortage of ideal SPSs, we adopt the decoy-state method \cite{Decoy-Lo,Decoy-Wang} to investigate the performance of our protocol with weak coherent sources (WCSs). With the widely used 3-intensity decoy-state ($\mu $, $\nu $, $0$) scheme, we can estimate the upper (lower) bound of the transmission rates $Y_{j\alpha ',s\chi '}^{{\Psi ^ - },U}$ ($Y_{j\alpha ',s\chi '}^{{\Psi ^ - },L}$) by the linear programming method or analytical formula, and further estimate the upper (lower) bound of ${{\rm{q}}^{\rm{T}}}$ with Eq. (\ref{equ}).  Then, the upper bound of $ e_{\alpha \chi }^{11,U}$ can be written as
\begin{equation}\label{E-U}
	e_{\alpha \chi }^{11,U} = \frac{1}{{1 + \frac{{{{(Y_{0\alpha ,1\chi }^{{\Psi ^ - },{\rm{vir}}} + Y_{1\alpha ,0\chi }^{{\Psi ^ - },{\rm{vir}}})}^L}}}{{{{(Y_{0\alpha ,0\chi }^{{\Psi ^ - },{\rm{vir}}} + Y_{1\alpha ,1\chi }^{{\Psi ^ - },{\rm{vir}}})}^U}}}}},
\end{equation}
where ${{{(Y_{0\alpha ,1\chi }^{{\Psi ^ - },{\rm{vir}}} + Y_{1\alpha ,0\chi }^{{\Psi ^ - },{\rm{vir}}})}^L}}$ is bounded by the linear programming problem $\mathop {\min }\limits_{{{\rm{q}}^{\rm{T}}}} (P_{0\alpha }^{{\rm{vir}}}P_{1\chi }^{{\rm{vir}}}{\rm{P}}_{0\alpha ,1\chi }^{{\rm{vir}}}{{\rm{q}}^{\rm{T}}} + P_{1\alpha }^{{\rm{vir}}}P_{0\chi }^{{\rm{vir}}}{\rm{P}}_{1\alpha ,0\chi }^{{\rm{vir}}}{{\rm{q}}^{\rm{T}}}) $, and ${{{(Y_{0\alpha ,0\chi }^{{\Psi ^ - },{\rm{vir}}} + Y_{1\alpha ,1\chi }^{{\Psi ^ - },{\rm{vir}}})}^U}}$ is bounded by the linear programming problem $\mathop {\max }\limits_{{{\rm{q}}^{\rm{T}}}} (P_{0\alpha }^{{\rm{vir}}}P_{0\chi }^{{\rm{vir}}}{\rm{P}}_{0\alpha ,0\chi }^{{\rm{vir}}}{{\rm{q}}^{\rm{T}}} + P_{1\alpha }^{{\rm{vir}}}P_{1\chi }^{{\rm{vir}}}{\rm{P}}_{1\alpha ,1\chi }^{{\rm{vir}}}{{\rm{q}}^{\rm{T}}})$. Finally, the secret key rate of our protocol based on WCSs is given by
\begin{equation}\label{R-WCS}
	R = P_{\mu \mu }^{11}Y_{ZZ}^{11,L}[1 - {I_E}] - Q_{ZZ}^{\mu \mu }fH(E_{ZZ}^{\mu \mu }),
\end{equation}
where $P_{\mu \mu }^{11}$ is the joint probability of both Alice's and Bob's signal states $\mu $  emitting single photon pulses, $Y_{ZZ}^{11,L} $ is the lower bound of single-photon counting rate in $ZZ$ bases, and $Q_{ZZ}^{\mu \mu }$ ($E_{ZZ}^{\mu \mu }$) is the overall gain (QBER) of signal states in $ZZ$.

\section{Numerical simulations}\label{4}

In this section, we present the simulation results of our protocol. Assume the four states prepared by Alice (Bob) are ${\left| {{\phi _{0Z}}} \right\rangle } = \cos \left( {\frac{{\delta _1  }}{2}} \right)\left| {{0_Z}} \right\rangle  + \sin \left( {\frac{{\delta _1 }}{2}} \right)\left| {{1_Z}} \right\rangle $, ${\left| {{\phi _{1Z}}} \right\rangle } = \sin \left( {\frac{{\delta _2  }}{2}} \right)\left| {{0_Z}} \right\rangle  + \cos \left( {\frac{{\delta _2  }}{2}} \right)\left| {{1_Z}} \right\rangle$, ${\left| {{\phi _{0X}}} \right\rangle } = \sin \left( {\frac{\pi }{4} + \frac{{\delta _3  }}{2}} \right)\left| {{0_Z}} \right\rangle  + \cos \left( {\frac{\pi }{4} + \frac{{\delta _3  }}{2}} \right){e^{i(\theta _1  + {\beta   })}}\left| {{1_Z}} \right\rangle $ and ${\left| {{\phi _{0Y}}} \right\rangle } = \sin \left( {\frac{\pi }{4} + \frac{{\delta _4  }}{2}} \right)\left| {{0_Z}} \right\rangle  + \cos \left( {\frac{\pi }{4} + \frac{{\delta _4  }}{2}} \right){e^{i\left( {\frac{\pi }{2} + \theta _2   + {\beta  }} \right)}}\left| {{1_Z}} \right\rangle$, where $\{ \delta _1 ,\delta _2  ,\delta _3  ,\delta _4  \} $ denote the flaws of time-bin  and $\{ \theta _1  ,\theta _2  \} $ are the imperfections introduced by PM. Note that the above state require a complete characterization. In practice, they can be estimated from a set of experimental data that minimize the key rate \cite {QKD-sourceflaw,BB84-exper-sourceflaw,MDI-exper-sourceflaw,RFI-exper2}. Here, we simply assume ${\theta _1}{\rm{ = }}{\theta _2}$ which can be attributed to the rotation angle $\beta $. Since RFI-MDI-QKD is insensitive to reference frames drift, we assume $\beta  = 0$ in the simulation below for simplicity. The detection efficiency and dark count rate of single-photon detectors are 14.5\% and $6.02 \times {10^{ - 6}}$ \cite{MDI2}, and the loss coefficient of the channel is 0.2 dB/km. 

\begin{figure}[htb]
	\centering
	\includegraphics[scale=0.5]{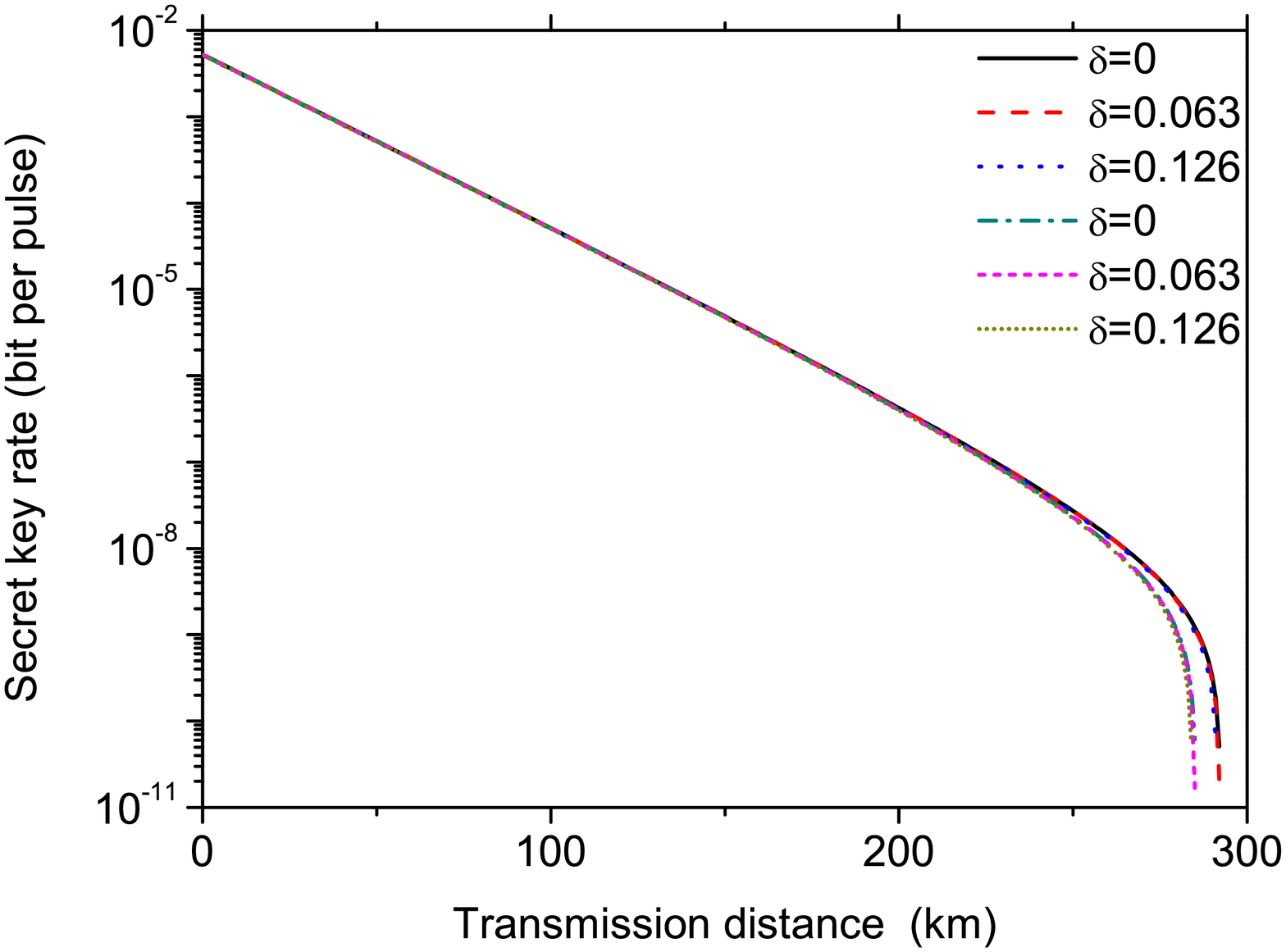}
	\caption{Secret key rates of the loss-tolerant RFI-MDI-QKD and MDI-QKD based on SPSs. The solid, dash and dot curves correspond to the secure key rates of the loss-tolerant RFI-MDI-QKD when $\delta {\rm{ = }}0,0.063{\rm{,0}}{\rm{.126}}$ \cite{source-error-1,source-error-2}, respectively. The dash dot, short dash and short dot curves correspond to the secure key rates of the loss-tolerant MDI-QKD when $\delta {\rm{ = }}0,0.063{\rm{,0}}{\rm{.126}}$, respectively.}
	\label{fig:Fig2}
\end{figure}

Fig. \ref{fig:Fig2} shows the secret key rates of the loss-tolerant RFI-MDI-QKD based on SPSs, where we assume ${\delta _1}{\rm{ = }}{\delta _2}{\rm{ = }}{\delta _3}{\rm{ = }}{\delta _4}{\rm{ = }}\delta $. For comparison, we also plot the the secret key rates of loss-tolerant MDI-QKD \cite{MDI-exper-sourceflaw} based on SPSs in Fig. \ref{fig:Fig2}. As for loss-tolerant MDI-QKD protocol, we set the probability of the emitted actual states as 1/3, ${\delta _1}{\rm{ = }}{\delta _2}{\rm{ = }}{\delta _3}{\rm{ = }}\delta $. As shown in Fig. \ref{fig:Fig2}, the secret key rates corresponding to the cases $\delta {\rm{ = }}0,0.063{\rm{,0}}{\rm{.126}}$ are almost overlap, which exhibit that both loss-tolerant RFI-MDI-QKD and MDI-QKD are robust against source flaws, and the performance of loss-tolerant RFI-MDI-QKD is better than that of loss-tolerant MDI-QKD, which is mostly due to the fact that Eve’s information is estimated more accurately in RFI-MDI-QKD than MDI-QKD. Hence, RFI-MDI-QKD is more suitable for scenarios when the alignment of reference frames is complicated.

\begin{figure}[htb]
	\centering
	\includegraphics[scale=0.5]{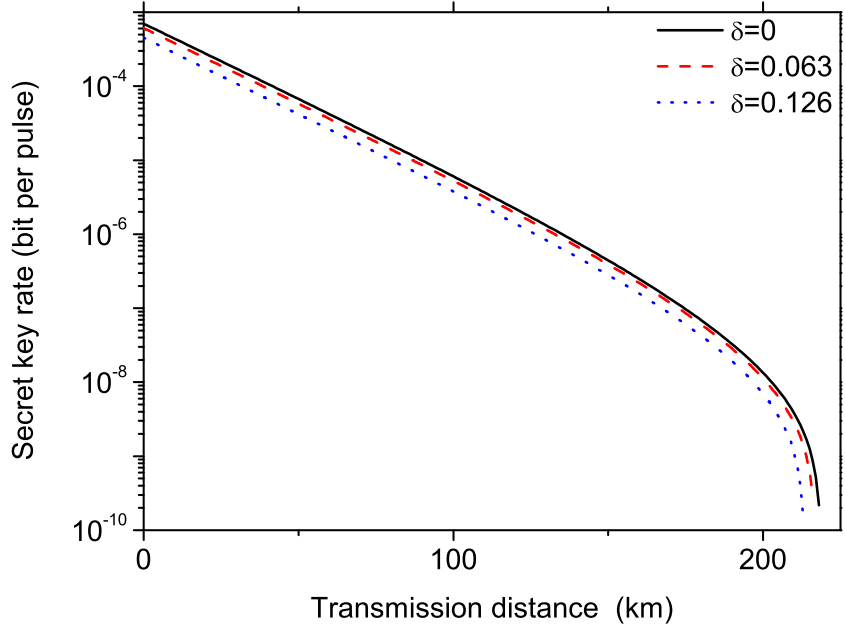}
	\caption{Secret key rates of the proposed RFI-MDI-QKD based on WCSs. The solid, dash and dot curves correspond to the secure key rates of our scheme when $\delta {\rm{ = }}0,0.063{\rm{,0}}{\rm{.126}}$, respectively.}
	\label{fig:Fig3}
\end{figure}

\begin{figure}[htb]
	\centering
	\includegraphics[scale=0.5]{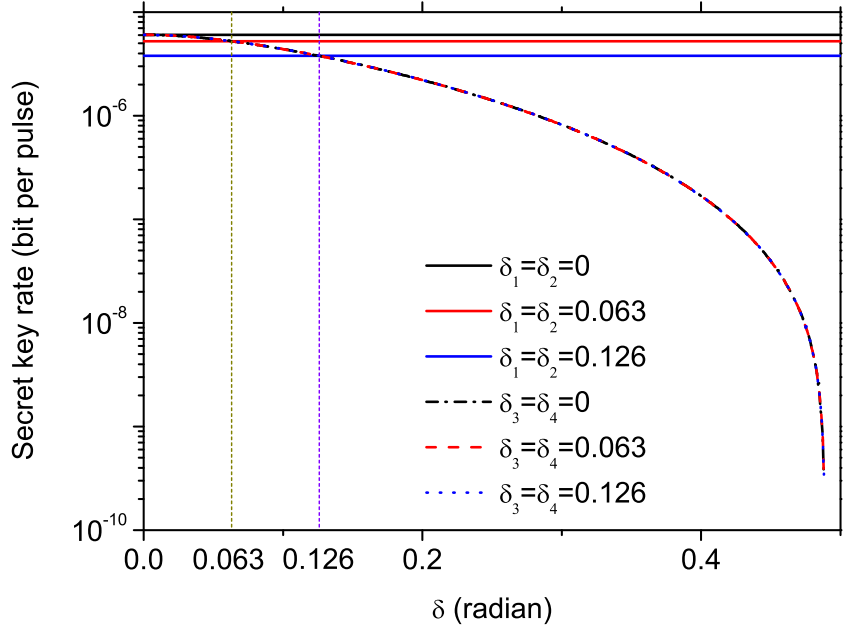}
	\caption{Secret key rates of decoy-state RFI-MDI-QKD versus the source flaws in different bases at 100 km. The three solid lines from top to bottom refer to the secret key rates versus the source flaws in $X$ and $Y$ bases (${\delta _{\rm{3}}}{\rm{ = }}{\delta _{\rm{4}}}{\rm{ = }}\delta $) with the source flaws in $Z$ basis fixed as ${\delta _1}{\rm{ = }}{\delta _2}{\rm{ = }}0,0.063,0.126$, respectively. And the three dashed to dotted curves refer to the secret key rates versus the source flaws in $Z$ basis (${\delta _{\rm{1}}}{\rm{ = }}{\delta _{\rm{2}}}{\rm{ = }}\delta $) with the source flaws in $X$ and $Y$ bases fixed as ${\delta _3}{\rm{ = }}{\delta _4}{\rm{ = }}0,0.063,0.126$, respectively.}
	\label{fig:Fig4}
\end{figure}

\begin{figure}[htb]
	\centering
	\includegraphics[scale=0.5]{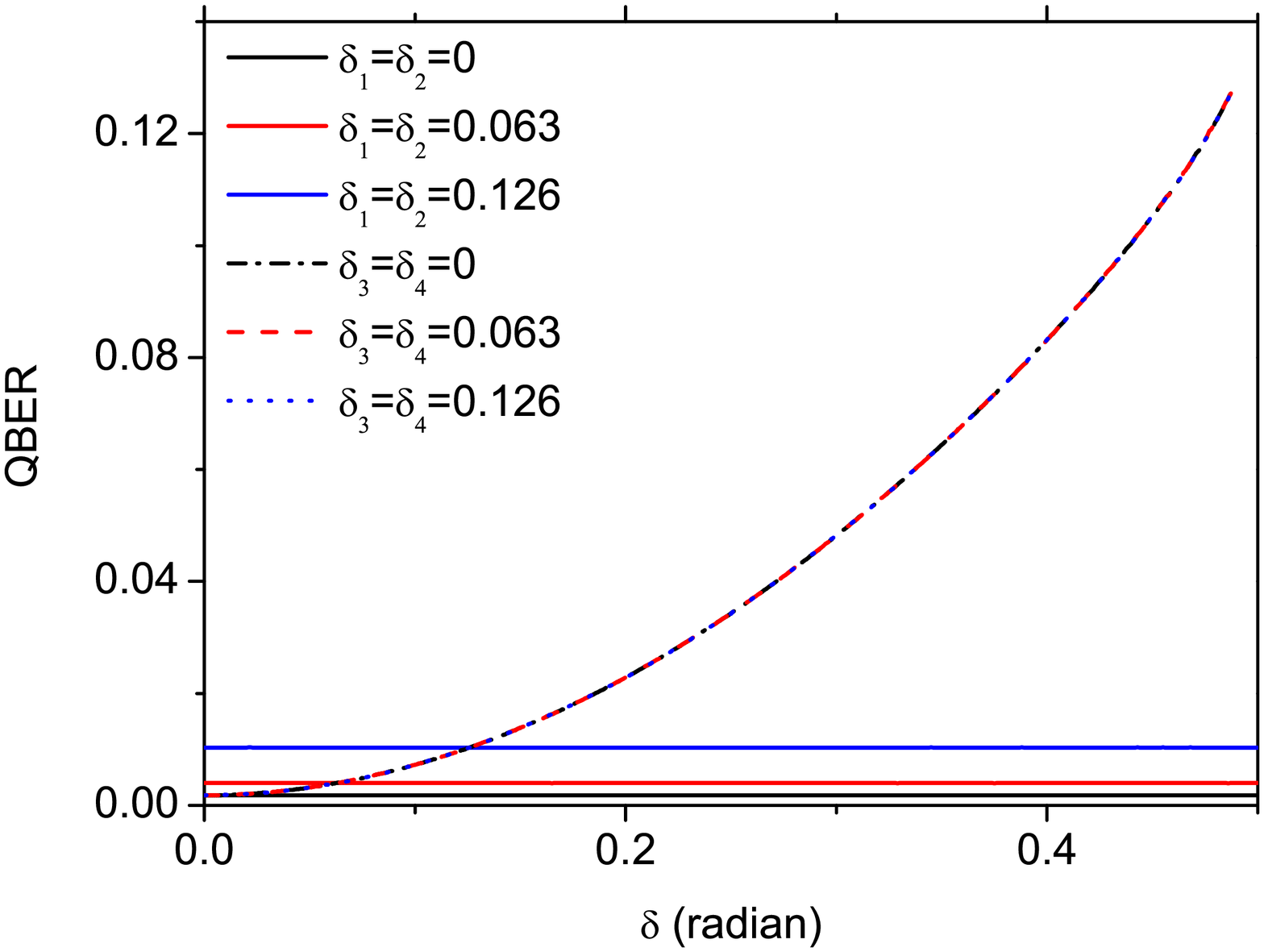}
	\caption{QBER of decoy-state RFI-MDI-QKD versus the source flaws in different bases at 100 km. The three solid lines from bottom to top refer to QBER versus the source flaws in $X$ and $Y$ bases (${\delta _{\rm{3}}}{\rm{ = }}{\delta _{\rm{4}}}{\rm{ = }}\delta $) with the source flaws in $Z$ basis fixed as ${\delta _1}{\rm{ = }}{\delta _2}{\rm{ = }}0,0.063,0.126$, respectively. And the three dashed to dotted curves refer to QBER versus the source flaws in $Z$ basis (${\delta _{\rm{1}}}{\rm{ = }}{\delta _{\rm{2}}}{\rm{ = }}\delta $) with the source flaws in $X$ and $Y$ bases fixed as ${\delta _3}{\rm{ = }}{\delta _4}{\rm{ = }}0,0.063,0.126$, respectively.}
	\label{fig:Fig5}
\end{figure}

 We also investigate the performance of loss-tolerant RFI-MDI-QKD based on WCSs with optimized intensities. In Fig. \ref{fig:Fig3}, the solid, dash and dot curves correspond to the cases $\delta {\rm{ = }}0,0.063{\rm{,0}}{\rm{.126}}$, respectively. Although the secret key rate slightly declines with the increase of $\delta $, a secure transmission distance of more than 240 km can still be achieved even with a relatively large encoding error. 
 
 In Fig. \ref{fig:Fig4}, we investigate the effect of source flaws in different bases on the secret key rates of decoy-state RFI-MDI-QKD at 100 km. The three solid lines from top to bottom refer to the secret key rates versus the source flaws in $X$ and $Y$ bases (${\delta _{\rm{3}}}{\rm{ = }}{\delta _{\rm{4}}}{\rm{ = }}\delta $) with the source flaws in $Z$ basis fixed as ${\delta _1}{\rm{ = }}{\delta _2}{\rm{ = }}0,0.063,0.126$, respectively. And the three curves refer to the secret key rates versus the source flaws in $Z$ basis (${\delta _{\rm{1}}}{\rm{ = }}{\delta _{\rm{2}}}{\rm{ = }}\delta $) with the source flaws in $X$ and $Y$ bases fixed as ${\delta _3}{\rm{ = }}{\delta _4}{\rm{ = }}0,0.063,0.126$, respectively. It can be seen that the three curves are almost overlap and decline dramatically with the increase of source flaws $\delta $ in $Z$ basis. We also find that the three solid lines remain nearly constant with respect to the source flaws $\delta $ in $X$ and $Y$ bases, and in this case, the difference between these three solid lines is determined by the source flaws $\delta $ in $Z$ basis. Besides, the horizontal values of the intersections of curves and solid lines are $\delta  = 0,0.063,0.126$, respectively. As illustrated in Fig. \ref{fig:Fig4}, the source flaws in $Z$ basis have adverse effect on RFI-MDI-QKD while the source flaws $\delta $ in $X$ and $Y$ bases have almost no effect on the key rates. Similar results were demonstrated in Ref. \cite{RFI-source}. In Fig. \ref{fig:Fig5}, we investigate the effect of source flaws in different bases on the QBER of decoy-state RFI-MDI-QKD at 100 km, as the QBER plays an important role in the error correction step. As shown in Fig. \ref{fig:Fig5}, source flaws in $X$ and $Y$ bases almost have no effect on the QBER, while the QBER grows with the increase of source flaws $\delta $ in $Z$ basis. Therefore, it is important to prepare $Z$-basis states accurately in RFI-MDI-QKD so as to generate more secret keys, and the $X$ and $Y$ bases states not as accurate as they should be, which will simplify experimental implementations of RFI-MDI-QKD.  

\begin{figure}[htb]
	\centering
	\includegraphics[scale=0.45]{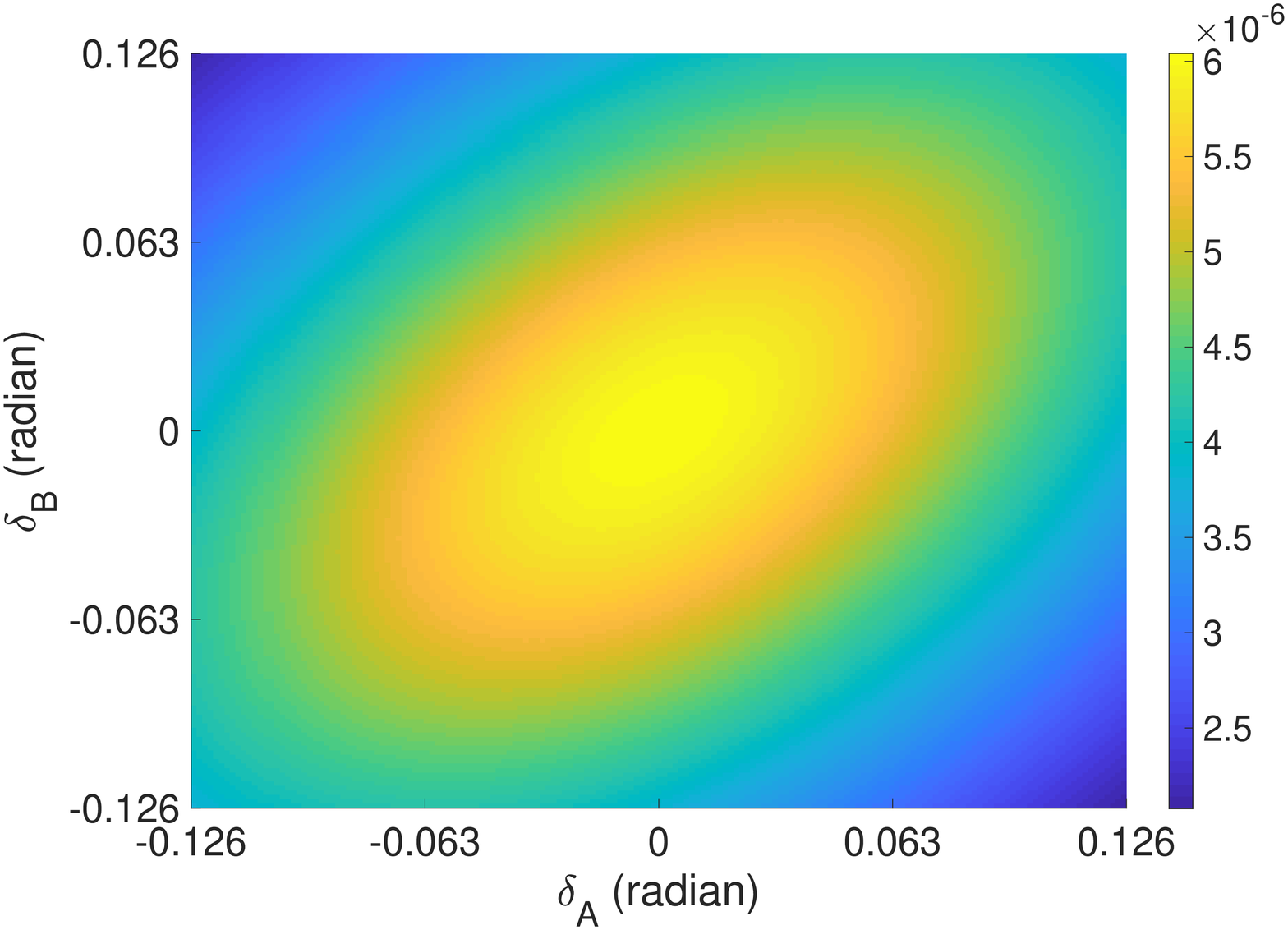}
	\caption{Secret key rates of decoy-state RFI-MDI-QKD as a function of Alice's and Bob's source flaws in $Z$ basis at 100 km.}
	\label{fig:Fig6}
\end{figure}

Since the flaws in $Z$ basis have obvious adverse effect on our protocol, we thoroughly investigate how Alice's and Bob's source flaws in $Z$ basis affect the secret key rates. Due to the trivial effect of source flaws in $X$ and $Y$ bases, we set their corresponding flaws as zero here. Denote Alice's source flaws in $Z$ basis  as ${\delta _1}{\rm{ = }}{\delta _2}{\rm{ = }}{\delta _A}$ and Bob's source flaws in $Z$ basis  as ${\delta _1}{\rm{ = }}{\delta _2}{\rm{ = }}{\delta _B}$. We show the secret key rates of decoy-state RFI-MDI-QKD as a function of ${\delta _A}$ and ${\delta _B}$ at 100 km in Fig. \ref{fig:Fig5}, which provides a vivid illustration when Alice's and Bob's source flaws in $Z$ basis are different.

\section{Conclusion}\label{5}

In conclusion, we have proposed the RFI-MDI-QKD protocol with imperfect sources. Comparing to the conventional RFI-MDI-QKD which needs six encoding quantum states, our protocol only prepares four states, which simplifies the implementation of RFI-MDI-QKD. Our simulation results indicate that the $Z$-basis states should be accurately prepared to generate higher secret key rates, while other states can tolerate relatively large flaws so as to simplify experimental implementations. Particularly, our protocol is suitable for QKD networks due to the various imperfect sources and misalignments of reference frames inherently existed among multiple senders, which will improve the practical security and reduce the implementation complexity in QKD networks. Besides, with the statistical fluctuation analysis in the future, the security of our protocol can be further improved.
 
\medskip

{\it Acknowledgments.} This work was supported by the National Key Research and Development Program of China (Grant Nos. 2018YFA0306400, 2017YFA0304100), the National Natural Science Foundation of China (Grant Nos. 61705110, 61590932, 11774180),  the China Postdoctoral Science Foundation (Grant Nos. 2019T120446, 2018M642281), and the Jiangsu Planned Projects for Postdoctoral Research Funds (Grant No. 2018K185C).

\bibliographystyle{IEEEtran}
\bibliography{IEEEabrv,Reference}

\end{document}